\newcommand{\be}{\begin{equation}}
\newcommand{\ee}{\end{equation}}
\documentclass[twocolumn,showpacs,superscriptaddress,amsmath,amssymb]{revtex4}
\usepackage{graphicx}
\usepackage{dcolumn}
\usepackage{bm}
\usepackage[colorlinks]{hyperref}

\begin{document}
\title{Comment on "Spatial optical solitons in highly nonlocal
media" and related papers\\}

\author{Milan S. Petrovi\'c}
\affiliation{Institute of Physics, P.O.Box 57, 11001 Belgrade,
Serbia} \affiliation{Texas A\&M University at Qatar, P.O.Box 23874,
Doha, Qatar}

\author{Najdan B. Aleksi\'c}
\affiliation{Texas A\&M University at Qatar, P.O.Box 23874, Doha, Qatar}
\affiliation{Institute of Physics, University of Belgrade,
P.O.Box 68, 11080 Belgrade, Serbia}

\author{Branislav N. Aleksi\'c}
\affiliation{Institute of Physics, University of Belgrade, P.O.Box
68, 11080 Belgrade, Serbia} \affiliation{Weill Cornell Medicine -
Qatar, Qatar Foundation - Education City, P.O. Box 24144, Doha,
Qatar}

\author{Aleksandra I. Strini\'c}
\affiliation{Institute of Physics, University of Belgrade,
P.O.Box 68, 11080 Belgrade, Serbia}

\author{Milivoj R. Beli\'c}
\affiliation{Texas A\&M University at Qatar, P.O.Box 23874, Doha,
Qatar}

\date{October 18, 2016}
\begin{abstract} In a recent paper [A. Alberucci, C. Jisha, N.
Smyth, and G. Assanto, Phys. Rev. A \textbf{91}, 013841 (2015)],
Alberucci \textit{et al.} have studied the propagation of bright
spatial solitary waves in highly nonlocal media. We find that the main results in that and related papers, concerning soliton shape and dynamics, based on the accessible soliton (AS) approximation,
are incorrect; the correct results have already been published by others.
These and other inconsistencies in the paper follow from the problems in applying the AS approximation in earlier papers by the group that propagated to the later papers. The accessible soliton theory cannot describe accurately the features and dynamics of solitons in highly nonlocal media.
\end{abstract}
\maketitle

Snyder and Mitchell introduced in 1997 a model of nonlinearity whose
response is highly nonlocal \cite{snyder}. They proposed an
elegant theoretical model, intimately connected with the linear
harmonic oscillator that described complex soliton dynamics in
simple terms.
Because of the simplicity of the theory, they coined the
term "accessible solitons" (ASs) for these optical spatial solitary
waves.

An early experimental observation of accessible solitons was reported in \cite{conti1,conti2}:
"We believe that most of the observed spatial optical solitons in
nematic liquid crystals (NLCs) are indeed accessible solitons,
inasmuch as NLC are highly nonlocal." The same authors have determined the basic beam evolution
laws for highly nonlocal NLCs in \cite{conti1,conti2}, which are later elaborated in \cite{alber1,alber2}. Equation (6) in \cite{conti2} was used to
interpret the experiments; the authors claimed good
agreement between the data and model predictions.

However, straightforward application of the AS approximation, even in
nonlinear media with almost infinite range of nonlocality,
inevitably leads to additional problems \cite{conti1,conti2,hennin},
because there exists no real physical medium without boundaries and
without losses. To include the impact of the finite size of the sample,
we developed a variational approach (VA) to solitons in nonlinear
media with long-range nonlocality, such as NLCs \cite{najdan1}.
Our VA results are corroborated by numerical simulations, and even have invited a comment
\cite{assanto1,najdan2}.

We highlighted the differences between accessible soliton
approximation and variational approach in nonlocal nonlinear media
in \cite{banePS}. The major differences are linked to the soliton
shape and dynamics:
\be R_{VA} = \sqrt{2} R_{AS}   , \quad \quad  \Lambda_{VA} = 2
\Lambda_{AS} , \label{RL_as_va} \ee

\noindent where $R$ is the beam width and $\Lambda$ is the period of
small oscillations around the equilibrium. Thus, in crucial characteristics,
the AS approximation is an oversimplification that at best can only
qualitatively apply to solitons in highly nonlocal media. The first published accurate quantitative
correction to AS approximation was presented in \cite{najdan1},
direct comparison between AS theory and VA was given in
\cite{banePS}, while a complete comparison in D=1,2,3 spatial
dimensions can be found in \cite{baneOE}.

In \cite{alber1}, published after our papers \cite{najdan1}
and \cite{banePS}, Alberucci \textit{et al.} discussed the main
features of ASs in realistic diffusive self-focusing
media. Authors concluded that "the
highly nonlocal (AS) model does not accurately describe the soliton
and predicts for it a width \textit{roughly} $\sqrt{2}$ smaller than
the actual size. This discrepancy stems from the role of the
boundary conditions." They cited this result again in
\cite{alber2} (this time as "discrepancy is due to the
singularity (at the origin) of the response function used here") and again, as their own.
But this crucial result was published before in
\cite{najdan1} and \cite{banePS}. None of the papers \cite{najdan1,baneOE,banePS}, where these and other related matters were discussed priorly, were cited in any of the papers by the above group.
In addition, some questions remain unanswered, concerning recent work by the group.

First, it is unclear why AS approximation cannot deal with highly nonlocal
situations, i.e. what is the reason for the appearance of \textit{rough} factors
$\sqrt{2}$ and 2 in the expressions for soliton shape and
dynamics: is it boundary conditions \cite{alber1} or singularity of
the response function \cite{alber2}?

Next, it remains unclear why authors omitted to cite accurate results for the beam
width and the period of small oscillations presented in
\cite{najdan1}, although they wrote and published a comment
\cite{assanto1} on that paper.
Instead, in \cite{alber1} they chose to cite the paper by Ouyang {\it et al.} \cite{ouyang2}, and in \cite{alber2} they cited themselves. However, the relevance of \cite{ouyang2} to the problems at hand is indirect;
it analyzes the approximate solutions of strongly nonlocal solitons. It represents a perturbation to the AS model, and it does {\it not} mention $\sqrt{2}$ and 2 corrections.
In fact, these corrections cannot be obtained by the method used in \cite{ouyang2}.

Third, the same group considers the soliton shape and dynamics in \cite{alber2} (pages 5 and 6). It is not difficult to show that
their results for the width and the breathing period of the soliton are
{\it still} wrong. The correct results are:
\be w =  \frac{2}{{k_0}} \sqrt{\frac{2 \pi}{\alpha n_0 P}} , \quad
\quad \Lambda = \frac{4 \sqrt{2} \pi^2}{k_0 \alpha P}   .
\label{correct} \ee
where $P$ is the beam power and $\alpha$ the absorption coefficient, according to the notation adopted in \cite{alber2}.
The problem stems from the incorrect analysis following Eqs. (15) and (16); the results of that analysis, represented in Fig. 7, are
wrong. For example, Fig. 7(c) represents the wrong formula for the width of the soliton.
Thus, the statement below Eqs. (15) and (16) that the width of solitons "is $\sqrt{2}$ larger than that stemming from the Snyder-Mitchell model" cannot be substantiated by this analysis and must have come from a different source. Curiously, Fig. 7 also appears as Figs. 1 and 2 in \cite{alber1}, although the starting equations in the two papers are different!

There are other conceptual problems in that paper from which these
and other inconsistencies follow. For example, in Eqs. (1) and (2)
for the wave amplitude $A$ and the refractive index change $\phi$,
an inconsistent approximation is utilized: the second-order
derivative in $z$ in Eq. (1) is approximated by a first-order
derivative, whereas in Eq. (2) it is kept second-order. While such a
treatment is commonly used in the paraxial approximation to the
slowly-varying wave amplitude, it is may not be appropriate for
nonlocal systems of coupled equations. Thus, when $\phi$ is fast
changing in $z$ on the same scale as $A$, as it is in \cite{alber2},
that change might spoil the paraxial approximation for $A$. In
addition, when one utilizes the material equation (2) as a source of
nonlocality, that equation will also couple to the backward
propagating wave. The fast change in $\phi$ will then generate the
back-scattered wave, even when there is no input wave. The omission
of back reflection is another source of inconsistency in
\cite{alber2}.

Finally, an intriguing question still hangs over the claimed good
agreement between experimental data and AS approximation, reported
in Fig. 3 from \cite{conti2}: "There is good agreement with the
calculation, and the standard deviation is 7\%." Eleven years later
the same group states that AS approximation is not correct
\cite{alber1,alber2}. The difference between the new and old
approximation is \textit{roughly} $\sqrt{2}$  and 2 for the soliton
width and period. Therefore, the close clustering of experimental
points in Fig. 3 about the solid lines that are "the best fits from
the theory" is bogus, since it suggests a close agreement between
experiment and AS theory, where there is none. In addition, the
claim in Fig. 3(D) that "the best fit from the theory" for
$\Lambda^{-2}$ as a function of power is a straight line, cannot be
true. So, the question is: Are the experimental data in
\cite{conti2} still only 7\% away from the AS approximation, as
claimed, or not?


In conclusion, the appearance of extra factors 
in the soliton existence equations is the consequence of
systematic errors in the AS approximation committed by the group in their early works. The corrections reported in later works are appropriated from others, without citation.
There exist conceptual problems in \cite{alber2} which raise questions about the correctness of the results obtained.
But, more importantly, the widespread belief that the AS
model can quantitatively explain beam propagation in highly nonlocal media is
unjustified. This model is just a linear approximation to a highly nonlocal nonlinear problem.


This work was supported by the Ministry of Science of the Republic
of Serbia under the projects OI 171033 and 171006, and by the NPRP
7-665-1-125 project of the Qatar National Research Fund (a member of
the Qatar Foundation).


\begin{thebibliography}{99}

\bibitem{snyder} A. W. Snyder and D. J. Mitchell, Accessible solitons, Science \textbf{276}, 1538 (1997).


\bibitem{conti1} C. Conti, M. Peccianti, and G. Assanto, Route to nonlocality
and observation of accessible solitons, Phys. Rev. Lett.
\textbf{91}, 073901 (2003).

\bibitem{conti2} C. Conti, M. Peccianti, and G. Assanto, Observation of optical
spatial solitons in a highly nonlocal medium, Phys. Rev. Lett.
\textbf{92}, 113902 (2004).

\bibitem{alber1} A. Alberucci, C. P. Jisha, and G. Assanto,
Accessible solitons in diffusive media, Opt. Lett. \textbf{39}, 4317
(2014).

\bibitem{alber2} A. Alberucci, C. P. Jisha, N. F. Smyth, and G. Assanto,
Spatial optical solitons in highly nonlocal media, Phys. Rev. A
\textbf{91}, 013841 (2015).

\bibitem{hennin} J. F. Henninot, J. F. Blach, and M. Warenghem, The investigation of an
electrically stabilized optical spatial soliton induced in a nematic
liquid crystal, J. Opt. A: Pure Appl. Opt. \textbf{10}, 085104
(2008).

\bibitem{najdan1} N. B. Aleksi\'{c}, M. S. Petrovi\'{c}, A. I. Strini\'{c},
and M. R. Beli\'{c}, Solitons in highly nonlocal nematic liquid
crystals: Variational approach, Phys. Rev. A \textbf{85}, 033826
(2012).

\bibitem{assanto1} G. Assanto and N. Smyth, Comment on "Solitons in
highly nonlocal nematic liquid crystals: Variational approach",
Phys. Rev. A \textbf{87}, 047801 (2013).

\bibitem{najdan2} N. B. Aleksi\'{c}, M. S. Petrovi\'{c}, A. I. Strini\'{c},
and M. R. Beli\'{c}, Reply to "Comment on 'Solitons in highly
nonlocal nematic liquid crystals: Variational approach' ", Phys.
Rev. A \textbf{87}, 047802 (2013).

\bibitem{banePS} B. N. Aleksi\'{c}, N. B. Aleksi\'{c}, M. S. Petrovi\'{c},
A. I. Strini\'{c}, and M. R. Beli\'{c}, Variational approach versus
accessible soliton approximation in nonlocal, nonlinear media, arXiv:1311.6840v2 (2013); Phys.
Scr. T \textbf{162}, 014003 (2014).

\bibitem{baneOE} B. N. Aleksi\'{c}, N. B. Aleksi\'{c}, M. S. Petrovi\'{c},
A. I. Strini\'{c}, and M. R. Beli\'{c}, Variational and accessible
soliton approximations to multidimensional solitons in highly
nonlocal nonlinear media, arXiv:1410.2580v1 (2014); Opt. Express \textbf{22}, 31842 (2014).


\bibitem{ouyang2} S. Ouyang and Q. Guo, (1+2)-dimensional strongly nonlocal solitons, Phys. Rev. A \textbf{76}, 053833 (2007).

\end{thebibliography}
\end{document}